\documentclass{optica-article}

\journal{opticajournal} 

\articletype{Research Article}

\usepackage{subcaption}
\usepackage{caption}
\usepackage{lineno}

\begin{document}

\title{Methane sensing in the mid-IR using short wave IR photon counting detectors via non-linear interferometry}

\author{Arthur C. Cardoso,\authormark{1, *} Jinghan Dong,\authormark{1} Haichen Zhou,\authormark{1}  Siddarth K. Joshi, \authormark{1} and John G. Rarity\authormark{1}}

\address{\authormark{1}Dept. of Electrical and Electronic Engineering, University of Bristol, Merchant Venturers Building,
Woodland Road, Bristol BS8 1UB, UK}

\email{\authormark{*}arthur.cardoso@bristol.ac.uk} 


\begin{abstract} 
We demonstrate a novel MIR methane sensor shifting measurement wavelength to SWIR (1.55$\mu$m) by using non-linear interferometry. The technique exploits the interference effects seen in three-wave mixing when pump, signal, and idler modes make a double pass through a nonlinear crystal. The method allows sensing at wavelengths where detectors are poor ($>$3$\mu$m) and detection at wavelengths where photon counting sensitivity can be achieved. In a first experimental demonstration, we measured a small methane concentration inside a gas cell with high precision. This interferometer can be built in a compact design for field operations and potentially enable the detection of low concentrations of methane at up to 100m range. Signal-to-noise ratio calculations show that the method can outperform existing short wavelength ($\sim$1.65$\mu$m) integrated path differential absorption direct sensing at high ($>$$10^{-4}$) non-linear gain.
\end{abstract}

\section{Introduction}
Methane is the second most important greenhouse gas in the Earth's atmosphere, its global warming potential is up to eighty-four times more than carbon dioxide in a twenty years period\cite{IPCC,Etminam_2016}, providing a huge contribution to the rise of temperature and related climate change. The concentration of this gas in the atmosphere has been rising since the beginning of the industrial age, and it is caused by anthropogenic and natural sources\cite{Blunier_1993,Etheridge_1998,Wuebbles_2002}. Agricultural, use and extraction of fossil fuels, biomass burning and decomposition of organic matter in wastes and landfills are the most important contributions among the human $\text{CH}_4$ sources\cite{Bousquet_2006}. Thus, monitoring and control of methane emissions is crucial to reduce global warming and mitigate climate changes\cite{Shindel_2012}.

Over the last decades, scientists have endeavored to develop new accurate methods to measure methane concentrations in the air. Many different experiments exploring optical methane sensors were performed using spaceborne \cite{Butz_2011,Jacob_2016}, airborne\cite{Golston_2017,Cossel_2017} and short range sensors\cite{Cezard_2020,Titchener_2022}. They approach miscellaneous techniques like Integrated Path Differential Absorption (IPDA), Differential Absorption Lidar (DIAL), Dual-comb spectroscopy (DCS), and spectral analysis of light among other methods. Modern short-range DIAL methane sensors\cite{Cezard_2020,Titchener_2022} are able to detect sources with high spatial and depth resolution with high sensitivity, being the state-of-art technology to identify small sources up to hundreds of meters away.

This paper exploits a novel method for short-range methane sensing adapting the concept of `imaging with undetected light' schemes, first exploited in \cite{Lemos_2014,Cardoso_2018}. This technique makes use of quantum correlated light sources in a nonlinear interferometer to collect phase and amplitude images of an object, by detecting light that never interacted with it. Recently, this methodology has been extended for several applications like spectroscopy\cite{Lee_2020,Chiara_2021}, microscopy\cite{Kviatkovsky_2020,Kviatkovsky_2022}, quantum holography\cite{Topfer_2022}, optical coherence tomography\cite{Machado_2020,Vanselow_2020} and terahertz sensing \cite{Kutas_2020}. Methane exhibits a series of strong narrow ($\sim$GHz) absorption lines in the mid-infrared (MIR), with up to one hundred times greater absorption than at short wave infrared (SWIR). However, MIR light detection is quite challenging due to thermal noise background and low-efficiency detectors. Here we introduce a `gas sensing with undetected light' scheme that enables us to probe the detailed gas absorption spectrum with narrowband laser at MIR wavelengths where methane has a high absorbance while detection is carried out at a wavelength where low-noise and efficient detectors are available. Our technique involving a narrowband laser allows high-resolution spectroscopy beyond that of Fourier transform spectroscopy based non-linear interferometry\cite{Chiara_2021}, where the resolution misses these fine details. Our method could enable long-range sensing for outdoor applications, as the sample doesn't need to be inside an ultra-stable interferometer (see approach in section \ref{sec:SNR}).

\section{Experimental implementation}\label{sec:expimp}

\subsection{Experimental setup}

\begin{figure}[h]
\captionsetup{font={footnotesize}}
   \centering\includegraphics[scale=0.8]{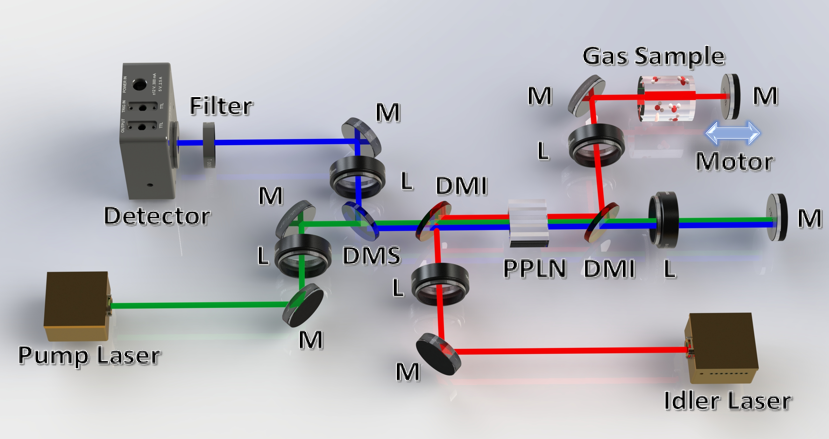}
      \caption{Experimental setup of the nonlinear interferometer used for sensing methane. A $1.064\mu$m and a tunable $3.22\mu$m laser interact with a PPLN crystal generating a signal beam by stimulated parametric down-conversion. The signal and the pump modes are reflected back to the crystal. The idler laser interacts with a gas cell and is reflected to the PPLN seeding a second down conversion. An interference pattern between the signal modes is measured by a photon-counting detector on shifting the position of the mirror that reflects the idler laser back to the crystal.}
      \label{fig:labsetup}
\end{figure}

The nonlinear interferometer proposed for methane sensing, depicted in Fig.\ref{fig:labsetup}, relies on (relatively) high-gain light sources generated by stimulated parametric down-conversion (PDC). A pump $1.064\mu$m continuous wave (CW) laser and an idler tunable $3.22\mu$m CW quantum cascade laser (Thorlabs ID3250HHLH) are overlapped in a nonlinear crystal (NLC), generating a signal mode around $1.59\mu$m by stimulated PDC. The pump and signal modes pass through a dichroic mirror (DMI) and are reflected back to the crystal. The idler field is directed by the DMI to pass through the gas cell and is then reflected to the NLC.  In the second pass through the crystal, the pump and idler mode generate a phase-dependent amplification of the signal field that is reflected to the detector by the dichroic mirror (DMS). A stepper motor shifts the position of the mirror that reflects the idler modes back to the crystal, adjusting the phase of the idler mode to produce amplification (constructive interference) or deamplification (destructive interference) at the second down conversion. As a result, the signal mode generated will show an interference pattern with period reflecting the idler mode wavelength. The pump laser power is $65$mW and the idler power is about $0.5$mW. We used a single photon counting detector of efficiency  $\approx 32\%$ efficiency and a neutral density filter of $\approx 2\%$ transmittance to keep the count rate under the saturation rate of the detector.

The intensity of the signal mode generated by the stimulated PDC is
\begin{equation}
     I_s = \left(\frac{\mu_0\epsilon_0\chi^{(2)}\omega_s^2L}{2k_s}\right)^2I_pI_i\\,
     \label{eq:intsignalgen}
\end{equation}
where $\mu_0$ is the magnetic vacuum permeability, $\epsilon_0$ is the electric vacuum permeability, $\chi^{(2)}$ is the effective non-linearity of the NLC, $L$ is its length, $\omega_s$ is the angular frequency of the signal mode, $k_s$ its wave-vector's modulus, $I_p$ the intensity of the pump laser and $I_i$ the intensity of the idler laser. Eq.(\ref{eq:intsignalgen}) can be written as $I_s = G I_i$, where $G =\left(\frac{\mu_0\epsilon_0\chi^{(2)}\omega_s^2L}{2k_s}\right)^2I_p$ is a gain factor related to the stimulated PDC process.
On the first pass through the crystal, we create a small amount of light $I_{s1}$ in the signal beam stimulated by the idler intensity $I_{i}$ and a small amount of extra light in the idler
\begin{equation}
    I_{s1} = GI_{i}, I_{i1}=I_{i}(1+G)\\.
    \label{eq:onepass}
\end{equation}
In our experiments with low-intensity CW pump light, $G<<1$ and we can ignore any higher-order (exponential) gain effects and assume no pump depletion. On returning through the crystal a second gain process occurs
\begin{equation}
    I_{s2} = GI_{i}\alpha\mathcal{T}^2\\,
    \label{eq:twopass}
\end{equation}
again ignoring higher-order terms in $G$. $\alpha$ is a term related to the losses 
in the optical components and spatial mode miss match and $\mathcal{T}$ is the transmission through the gas cell. 
The field generated in the first pass $E_{s1}$ returns to the crystal and interferes with the second pass $E_{s2}$ and the resulting intensity at the signal detector can be written as
\begin{equation}
    I_{s} = I_{s1}+I_{s2}+2\sqrt{I_{s1}I_{s2}}\cos\phi\\.
    \label{eq:intensig}
\end{equation}
 where 
 \begin{equation*}
  \phi=  \phi_{p}-\phi_{i}-\phi_{s}  
  \label{eq:phase}
 \end{equation*} 
is the relative phase between the pump, signal and idler fields, respectively. Changing the phase by translating any of the return mirrors will generate periodic fringe patterns with period given by the wavelength of the idler laser. The visibility $\mathcal{V}$ of this interference pattern contains a signature of the loss in the idler arm as 
\begin{equation}
    I_{s} = CGI_{i}(1+\mathcal{V}\cos\phi)\\.
    \label{eq:simpleinterf}
\end{equation}
where $C=1+\alpha\mathcal{T}^2$ varies from $[1-2]$, depending on loss in the idler arm and 
\begin{equation}
   \mathcal{V} = 2\mathcal{T}\sqrt{\alpha}/C\\.
    \label{eq:vis}
\end{equation}
Having determined the $\mathcal{T}$ from the visibility we can then evaluate methane concentration. The transmittance can be expressed as 
\begin{equation}
   \mathcal{T} = e^{-ZX_{\text{CH}_4}n_{\text{air}}\sigma}\\,
    \label{eq:Transmittance}
\end{equation}
where $Z$ is the length of the gas cell, $X_{\text{CH}_4}$ its methane concentration, $n_{\text{air}}$ is the density of air molecules, and $\sigma$ is the methane's absorption cross-section.

\subsection{Experimental results}

\begin{figure}[h]
\centering
\begin{minipage}{0.48\textwidth}
    \centering
    \captionsetup{margin=0cm}
    \includegraphics[width=\textwidth]{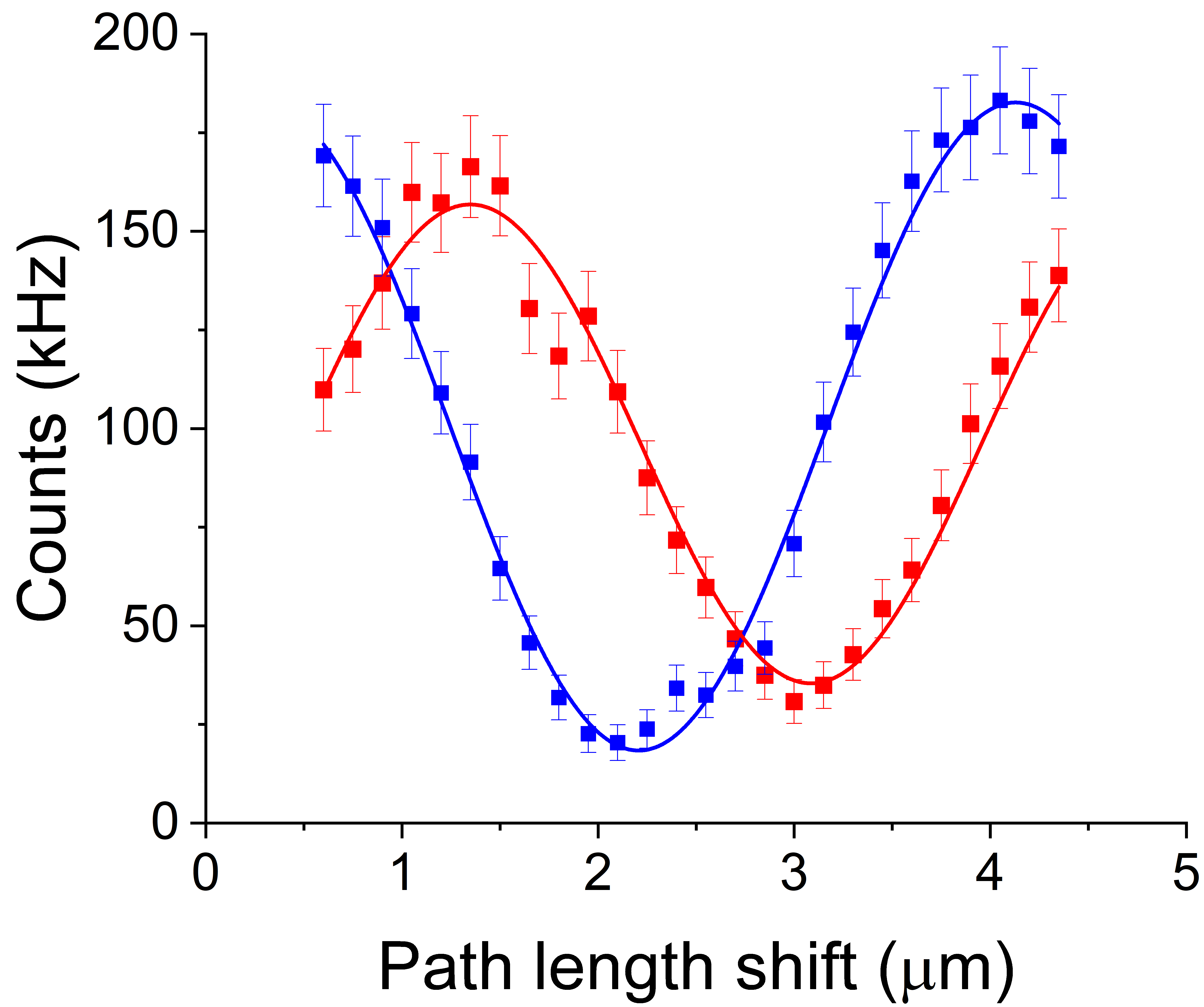}
    \captionof{figure}{Off- (blue dots) and on-resonance (red dots) interference patterns with sinusoidal line fits.}
    \label{fig:onandoffIP}
\end{minipage}%
\hfill
\begin{minipage}{0.48\textwidth}
    \centering
    \captionsetup{margin=0cm}
    \includegraphics[width=\textwidth]{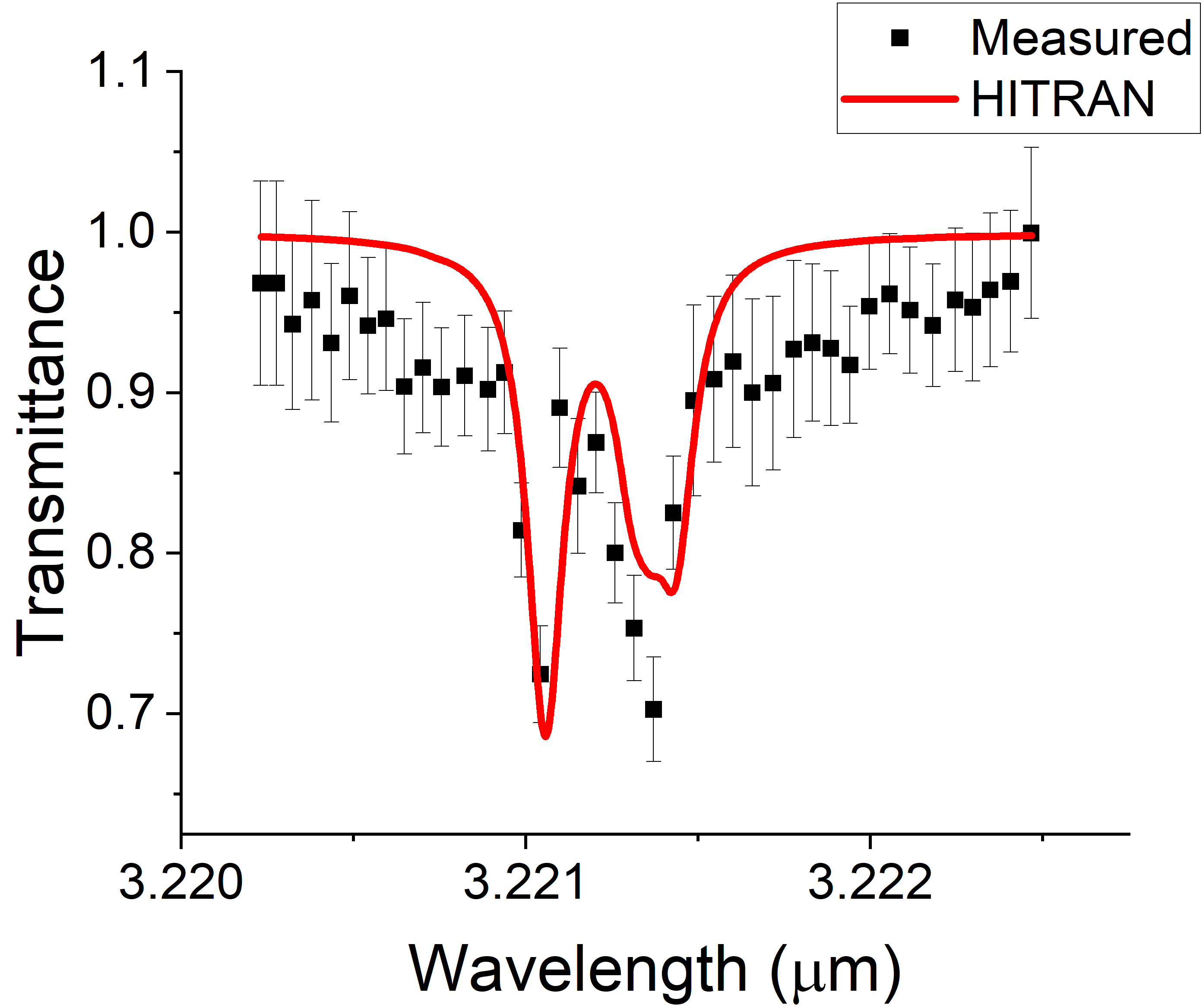}
    \captionof{figure}{Plots of transmittance as function of the wavelength calculated from the visibility data by using Eq.\ref{eq:vis} (black squares), and gas cell simulation data provided by HITRAN\cite{HITRAN} (red line).}
    \label{fig:transvswl}
\end{minipage}%
\label{fig:visandtransvswl}
\end{figure}

The apparatus was first aligned off-resonance to optimise the fringe visibility $\mathcal{V}$. Two example fringes (on- and off-resonance) measured by varying the idler mirror position are shown in Fig.\ref{fig:onandoffIP}. A full absorption spectrum was then collected by shifting the wavelength of the stimulating idler laser through the methane absorption feature sampling fringe visibility at a series of wavelengths. The visibilities of the on- and off-resonance interference patterns measured are $\mathcal{V}_{on}=\left(65.7 \pm 1.6\right)\%$ and $\mathcal{V}_{off}=\left(83.0 \pm 1.8\right)\%$, respectively, can be seen on Fig.\ref{fig:onandoffIP}. For the off-resonance wavelength, we consider $\mathcal{T}_{off}=1$ so that we can obtain the value of $\alpha = 0.28 \pm 0.02$ from Eq.\ref{eq:vis}. Then, using the same equation we calculate the transmittance of the gas cell $\mathcal{T}$ as function of the wavelength. The result, plotted in Fig.\ref{fig:transvswl}, exhibits a double peak absorption around $3.221\mu$m, similar feature to the one provided by HITRAN\cite{HITRAN}. The HITRAN data on Fig.\ref{fig:transvswl} shows the simulated transmittance of a methane gas cell of $2.5$cm at $294$K temperature, $5000$ppm concentration of methane, buffered to 1 atm pressure (with nitrogen gas $N_2$). This is the same specification as the gas cell used in the experiment.

The minimum transmittance obtained is $\mathcal{T}_{on}=0.70 \pm 0.03$. Using equation \ref{eq:Transmittance} we can calculate the methane concentration, inside the $2.5$cm gas cell. This yields a value $X_{\text{CH}_4}=\left(4779\pm 574\right)$ppm, by using $n_{\text{air}}=2.53 \times 10^{25}$ molec.m$^{-3}$ and the tabulated absorption cross-section \cite{HITRAN} of  $\sigma_{\text{on}} = 1.18 \times 10^{-22}\text{m}^2/\text{molec.}$ on-resonance. This transmittance value also includes the absorption of background methane throughout the interferometer, though this concentration is very low, usually between $1.5$ and $2.5$ppm.m. As our idler arm size is about $1$m the background methane absorption can be neglected in the measurements.

\section{Sensitivity evaluation and comparison with the direct sensing method}\label{sec:SNR}

\subsection{Signal-to-noise ratio}

The above first result is established in laboratory conditions with low loss $\alpha$ and artificially low detector efficiency of <1\% to prevent saturation. In this section we will analyse the sensitivity of the method in an outdoor and longer-range application scenario. In this case, the interferometer might be built using partial reflection from objects in the landscape or from suitably placed retro-reflection sites to return the idler. A phase shifter would be incorporated into the local setup to enable the measurement of constructive and destructive interference between the signal modes generated, also a narrow bandwidth ($\Delta\lambda\approx 1-2$nm) filter can be added to the detection to reduce the background noise from the sunlight. This apparatus could then be used to make stand-off measurements of methane inside a gas plume at distance (eg ~100m). In this longer range scenario, we can assume a relatively low return ($\alpha$<0.01) and make the assumption  $I_{s1}>>I_{s2}$ and treat this interference process in the same way as homodyne detection, identifying $I_{s1}$ with the local oscillator $I_{LO}$ and $I_{s2}$ with the signal $I_{sig}$ and written as
\begin{equation}
    I = I_{LO}+I_{sig}+2\sqrt{I_{LO}I_{sig}}\cos\phi\\.
    \label{eq:intendet}
\end{equation}

The signal-to-noise ratio (SNR) can then be analysed in an analogous way to homodyne interferometry \cite{Sun_2011}. The detected signal on- and off-resonance can be written in terms of the  difference between maximum and minimum photon counts of an interference fringe received, $R$, and its variance, $V$, 

\begin{equation}
    R = \frac{2\eta T_{int}}{\hbar\omega_s}\sqrt{\langle P_{sig}\rangle \langle P_{LO}\rangle}\\, \mbox{ and}
    \label{eq:signal}
\end{equation}

\begin{equation}
    V \approx \eta T_{int}\frac{\langle P_{LO}\rangle}{\hbar\omega_s}\\,
    \label{eq:noise}
\end{equation}
where: $\eta$ is the efficiency of the detector, $\hbar\omega_s$ is the energy of a signal photon $\langle P_{sig}\rangle$ is the power of the signal input, $\langle P_{LO}\rangle$ is the power of the local oscillator and $T_{int}$ is the measurement time. For comparison with direct detection schemes, we assume the signal is estimated from the difference of maximum and minimum of fringe measured in time $T_{int}/2$. Although the results shown in section \ref{sec:expimp} are not yet shot-noise limited, for simplification purposes, we will consider that this is the case now, that is, the noise produced by the photon number fluctuations is much greater than the environment and detector's noise. Also, as the noise provided by the signal is much smaller than the local oscillator's noise, we neglect it. So, the SNR on- and off-resonance can be written in terms of the number of photons combining Eq.(\ref{eq:signal}) and Eq.(\ref{eq:noise}) as

\begin{equation}
    SNR_{\text{on,off}}=\frac{R_{\text{on,off}}}{\sqrt{V_{\text{on,off}}}}=2\sqrt{ \langle n_{sig}\rangle_{\text{on,off}}}\\,
    \label{eq:SNRWD}
\end{equation}
where $\langle n_{sig}\rangle_{\text{on}}$ and $\langle n_{sig}\rangle_{\text{off}}$ are the average number of photons generated by the second stimulated PDC in the nonlinear crystal counted by the detector at on- and off- resonance, respectively
\begin{equation}
    \langle n_{sig}\rangle_{\text{on,off}}= \eta G\alpha\mathcal{T}^{2}_{\text{on,off}}T_{int}\frac{ P_i}{\hbar\omega_s}\\.
\end{equation}
where $\alpha$ is the reflectivity of the background taking into account the geometric losses, $\mathcal{T}_{\text{on}}$ and $\mathcal{T}_{\text{off}}$ are the transmittances of the gas plume for on- and off-resonance MIR idler wavelengths and $P_i$ is the power of the idler laser. In this case, we will take the background as a Lambertian or partially retroreflecting surface. Also, as the methane lines are narrow, we considered the same gain, detector efficiency, angular frequency and stimulating laser (idler) power for the on- and off-resonance frequencies.

\subsection{Sensitivity evaluation and comparison}

We use a similar analysis to that used to calculate the minimum detectable gas concentration in direct detection SWIR $CO_2$ IPDA Lidar in \cite{Quatrevalet_2017}. This allows us to make a direct comparison between the SWIR direct detection method and the MIR non-linear interferometry developed here. The transmittance of the gas plume can be written as
\begin{equation}
    \mathcal{T}_{\text{on,off}} = e^{-OD_{\text{on},\text{off}}} \\,
    \label{eq:TDAODMIR}
\end{equation}
with
\begin{equation}
    OD_{\text{on,off}} = Zn_{\text{CH}_4}\sigma_{\text{on,off}}  \\,
\end{equation}
where: $OD_{\text{on,off}}$ is the optical density and $\sigma_{\text{on,off}}$ is the effective absorption cross-section for on- and off-resonance MIR idler wavelength, and $n_{\text{CH}_4}$ and $Z$ are the number density of methane molecules and the depth of the gas plume, respectively. Thus, the differential absorption optical depth, $DAOD = OD_{\text{on}} - OD_{\text{off}}$ becomes
\begin{equation}
    DAOD  =  Zn_{\text{CH}_4}\left(\sigma_{\text{on}}-\sigma_{\text{off}}\right) = \ln\left(\frac{R_{\text{off}}}{R_{\text{on}}}\right)\ .
    \label{eq:DAOD}
\end{equation}
The quantity of scientific interest, the dry-air average volume mixing ratio of $\text{CH}_4$, $X_{\text{CH}_4}$, is related to $n_{\text{CH}_4}$ and number density of dry air $n_{\text{air}}$ via $X_{\text{CH}_4} = n_{\text{CH}_4}/n_{\text{air}}$. So, Eq.(\ref{eq:DAOD}) can be reformulated as
\begin{equation}
    X_{\text{CH}_4} =  \frac{DAOD}{Z n_{\text{air}}\left(\sigma_{\text{on}}-\sigma_{\text{off}}\right)}\\.
    \label{eq:XCH4}
\end{equation}

For a Gaussian approximation to Poisson noise, the methane detection sensitivity of the method can be estimated by Gaussian error propagation. Propagating the error in Eq.(\ref{eq:DAOD}), the detection precision can be written as
\begin{equation}
    \delta X_{\text{CH}_4} = \sqrt{\left(\frac{\partial X_{\text{CH}_4}}{\partial R_{\text{on}}}\right)^{2}V_{\text{on}}+\left(\frac{\partial X_{\text{CH}_4}}{\partial R_{\text{off}}}\right)^{2}V_{\text{off}}}
     = \frac{\sqrt{SNR^{-2}_{\text{on}}+SNR^{-2}_{\text{off}}}}{Zn_{\text{air}} \left(\sigma_{\text{on}}-\sigma_{\text{off}}\right)} \mbox{,}
\end{equation}
where $\delta X_{\text{CH}_4}$ is the minimum methane concentration that can be sensed by the apparatus. From Eq.(\ref{eq:SNRWD}), it follows that
\begin{equation}
    \delta X_{\text{CH}_4} =  \frac{1}{Zn_{\text{air}}\left(\sigma_{\text{on}}-\sigma_{\text{off}}\right)} 
    \sqrt{\frac{\hbar\omega_{s}}{2\eta G\alpha T_{int} P_i}}\mbox{.}
    \label{eq:deltaXCH4WD}
\end{equation}
As we are calculating the minimum amount of methane that can be detected by the apparatus, we expect a high transmittance for both on- and off-resonance idler frequencies, so we made $\mathcal{T}_{\text{on}}=\mathcal{T}_{\text{off}}\approx1$.

The SNR for a methane direct SWIR DIAL sensor \cite{Titchener_2022} can be obtained with a similar derivation as the one in \cite{Quatrevalet_2017}. In this case, the average number of photons detected on- and off-resonance is
\begin{equation}
    \langle n\rangle_{\text{on,off}}=\eta\alpha\mathcal{T}^{2}_{\text{on,off}}T_{int}\frac{P_{\text{on,off}}}{\hbar\omega_{\text{on,off}}}\mbox{,}
\end{equation}
where: $\eta$ is the efficiency of the detector, $\alpha$ is the reflectivity of the Lambertian background considering the geometric losses, $T_{int}$ is the measurement time, $\mathcal{T}_{\text{on,off}}$ is the transmittance, $P_{\text{on,off}}$ is the power of the laser and $\hbar\omega_{\text{on,off}}$ is the energy of a photon, for on- and off-resonance wavelengths. So, the SNR for the direct sensing in the shot-noise limited scenario is
\begin{equation}
    SNR=\sqrt{\langle n\rangle_{\text{on,off}}}=\mathcal{T}_{\text{on,off}}\sqrt{\eta\alpha T_{int}\frac{P_{\text{on,off}}}{\hbar\omega_{\text{on,off}}}}\mbox{.}
\end{equation}
It follows that the minimum amount of methane that can be detected by direct sensing is 
\begin{equation}
\begin{aligned}
    \delta X_{\text{CH}_4} = & \sqrt{\left(\frac{\partial X_{\text{CH}_4}}{\partial \langle n\rangle_{\text{on}}}\right)^{2}V_{\text{on}}+\left(\frac{\partial X_{\text{CH}_4}}{\partial \langle n\rangle_{\text{off}}}\right)^{2}V_{\text{off}}}\\
     = & \frac{1}{Zn_{\text{air}}\left(\sigma_{\text{on}}(1.65)-\sigma_{\text{off}}(1.65)\right)}
    \sqrt{\frac{\hbar\omega}{2\eta \alpha T_{int} P}}\mbox{,}
    \label{eq:deltaXCH4D}
\end{aligned}
\end{equation}
where $\sigma_{\text{on,off}}(1.65)$ is the effective absorption cross-section for on- and off-resonance SWIR laser wavelength, and $n_{\text{CH}_4}$ and $Z$ are the number density of methane molecules and the depth of the gas plume, respectively. We considered the same angular frequency and the same laser power for on- and off-resonance frequencies, also $\mathcal{T}_{\text{on}}=\mathcal{T}_{\text{off}}\approx1$.

We define the relative sensitivity between the two methods as,
\begin{equation}
    R_{S} = \frac{\delta X_{\text{CH}_4}\mbox{ for direct measurement}} {\delta X_{\text{CH}_4}\mbox{ for sensing with undetected light}}\ .
\end{equation}
This number provides us a quantitative comparison between both techniques. If we analyse a scenario that $R_{S} < 1$ it means that the direct sensing is more sensitive, if $R_{S} > 1$ the sensing with undetected light method is more sensitive. If we make an approximating assumption that detector efficiencies, reflectivity of the backscattering surface and measurement time are equal for both wavelength scenarios then taking the ratio of  Eq.(\ref{eq:deltaXCH4WD}) and Eq.(\ref{eq:deltaXCH4D}) leads to,
\begin{equation}
    R_{S} \simeq  \frac{\sigma_{\text{on}}(3.22)-\sigma_{\text{off}}(3.22)} {\sigma_{\text{on}}(1.65)-\sigma_{\text{off}}(1.65)}\sqrt{G}  \ .
\end{equation}
The values $\sigma$ can be obtained from HITRAN molecular spectroscopy database\cite{HITRAN}. In the case of the MIR line of $3.221\mu$m is around $65.1$ times greater than the SWIR one at $1.65\mu$m ($\sigma_{on}(1.65) = 1.81\times 10^{-24})$\cite{HITRAN}, so the sensitivity ratio is 

\begin{equation}
    R_S \simeq 65.1 \sqrt{G}  \ ,
    \label{eq:sensitivityratio}
\end{equation}
 where we considered $\sigma_{off}(1.65)\approx 0 $. The sensitivity ratio $R_S$ passes unity when $G>2.36\times 10^{-4}$ assuming our simplistic approximations hold (equal $P, \alpha, \eta, T_{int}$, etc.).  We assumed that the angular frequencies for the signal mode is the same as the one used in the direct sensing as well as the laser's power, i.e. $\omega_s=\omega$ and $P_{i}=P$. 
\subsection{Simulations}
\begin{figure}[h]
\centering
\begin{minipage}{0.48\textwidth}
    \centering
    \captionsetup{margin=0cm}
    \includegraphics[width=\textwidth]{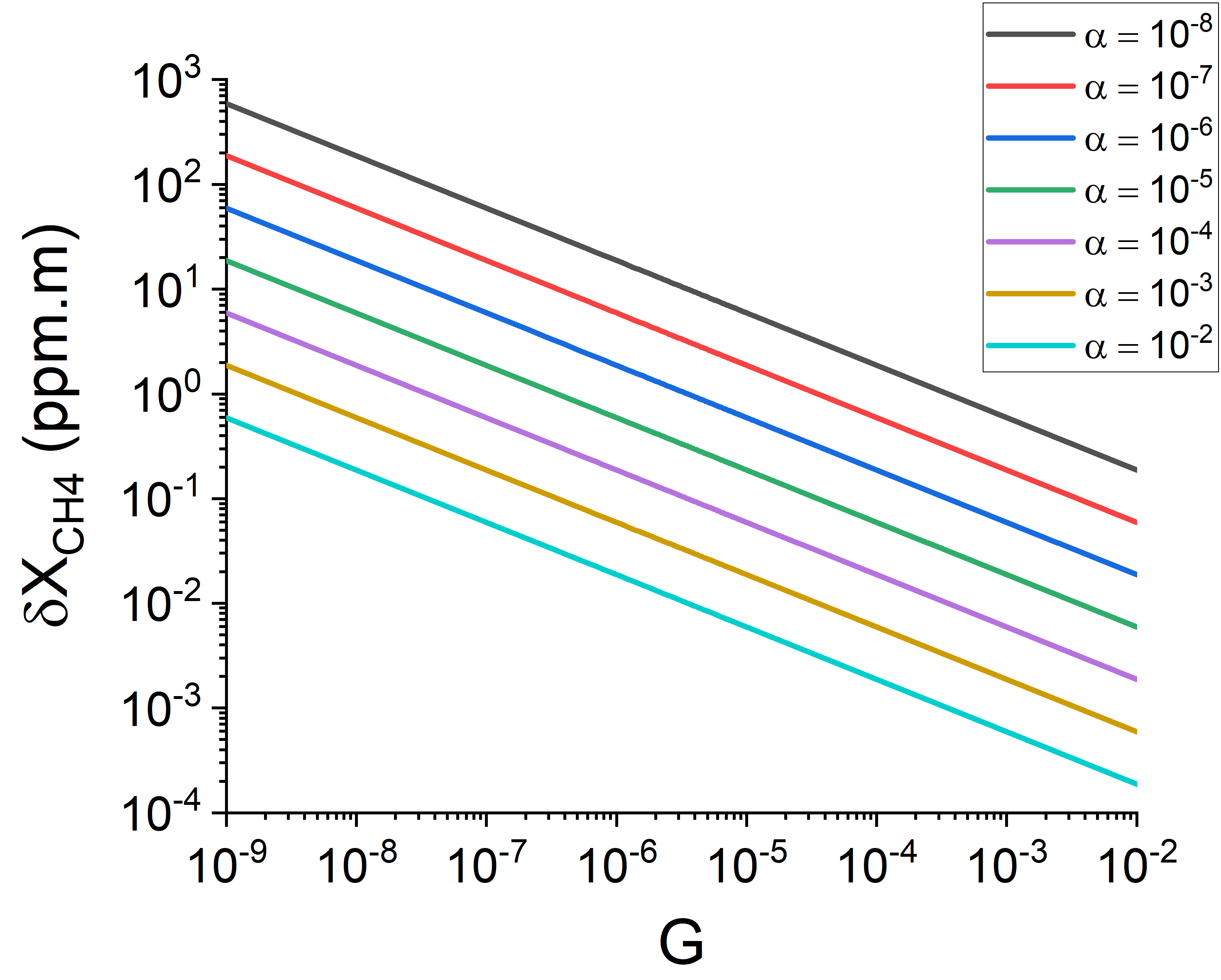}
    \captionof{figure}{The variation of $\delta X_{\text{CH}_4}$ (in ppm.m) with $G$ at different values of $\alpha$ shown on logarithmic scales.}
    \label{fig:XCH4-lgG}
\end{minipage}%
\hfill
\begin{minipage}{0.48\textwidth}
    \centering
    \captionsetup{margin=0cm}
    \includegraphics[width=\textwidth]{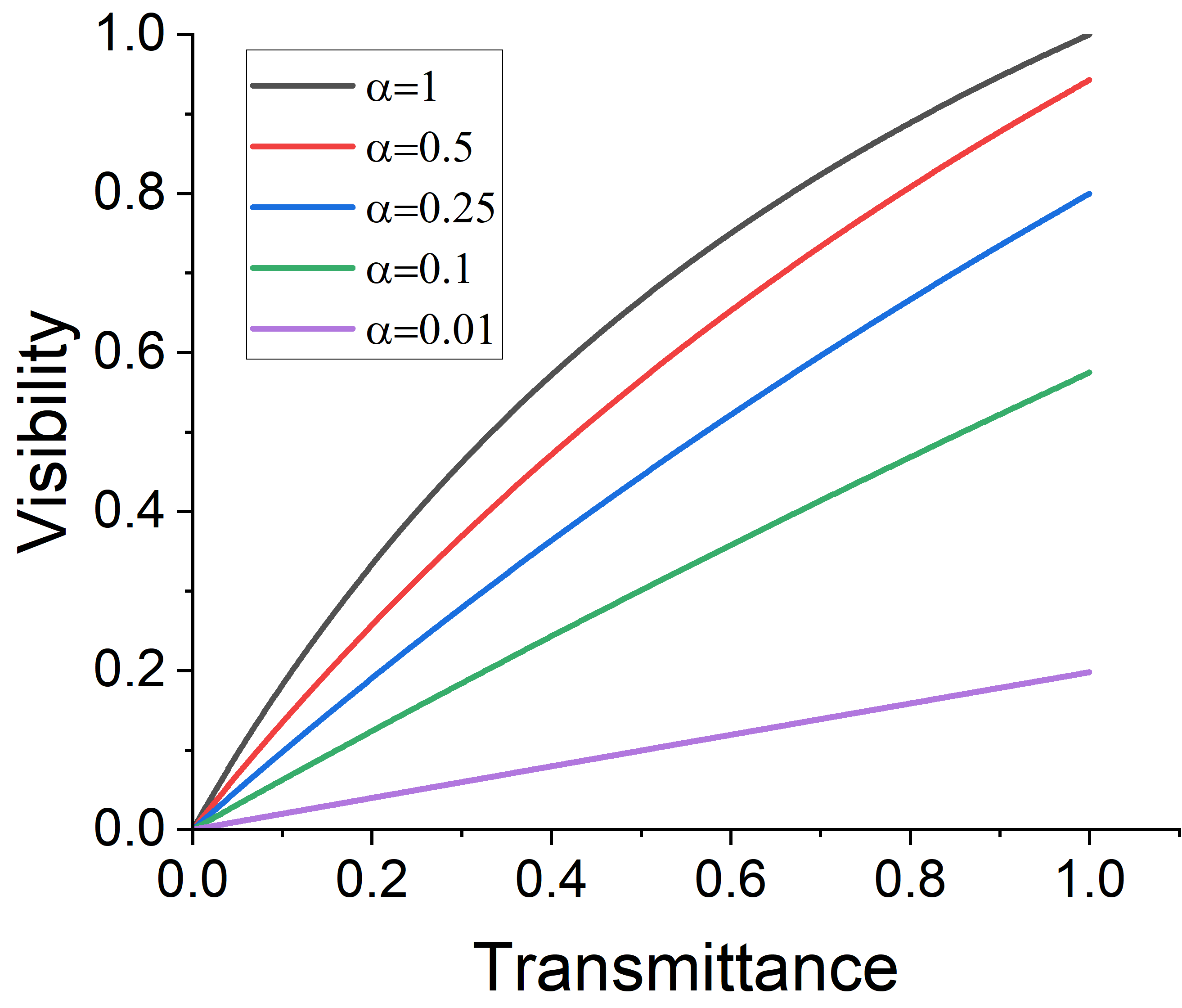}
    \captionof{figure}{Visibility as function of the gas cell's transmittance for different values of $\alpha$.}
    \label{fig:Visvstrans}
\end{minipage}
       
\label{fig:simulations}
\end{figure}

In a typical scenario for gas sensing we might expect parameters in Eq.(\ref{eq:deltaXCH4WD}) of the order
\begin{equation}
\begin{aligned}
    &\eta = 0.1\mbox{, } G = 10^{-8}\mbox{, }\alpha = 10^{-8} \mbox{, }  T_{int} = 1\text{s} \mbox{, } \\
     & P_i = 0.02 \text{W}\mbox{ and }
    \sigma_{\text{on}} = 1.18 \times 10^{-22}\frac{\text{m}^2}{\text{molec.}}  \,
    \label{eq:parametersdeltaXCH4}
\end{aligned}
\end{equation}
assuming the MIR line is the one in our experiment ($3.221\mu$m). We have considered a low-efficiency value for the detector in the simulation (10\%) reflecting our present experimental set-up but in the future, the high-intensity signal modes generated by stimulated PDC could be easily detected by a higher (80\%) efficiency InGaAs camera. Also, the gain value reflects the measured experimental gain which could be increased by reconfiguring experimental geometry with tighter focus, higher non-linearity or longer crystals. The value estimated for $\alpha$ is realistic if we consider a Lambertian background target about 100 meters away from the apparatus if a $100$mm diameter collection lens is used \cite{Titchener_2022}. With these parameters we can reach a sensitivity of $\delta X_{\text{CH}_4}\approx187$ppm.m, i.e., we can detect a methane cloud of $187$ppm concentration and one meter depth. This brings the sensitivity within the range required for gas leak detection. We then look at how this could be improved by increasing these conservative values of gain $G$ and return loss $\alpha$. Figure \ref{fig:XCH4-lgG} shows the variation of $\delta X_{\text{CH}_4}$ with $G$ for various values of return loss $\alpha$ and Figure \ref{fig:Visvstrans} the visibility as function of the transmittance for different values of $\alpha$. What is significant is that for short range scenarios where a retro-reflector is used and $\alpha>10^{-2}$ then the sensitivity is better than 1 ppm.m for all values of $G$ shown and reaches a few ppB.m at gain values $G>10^{-4}$. These sensitivity values are now potentially useful for monitoring small increases in environmental 'background' methane which is typically 1.5-2.5 ppm.m. 
When we consider the experimental results shown in Fig.\ref{fig:transvswl} we see a concentration $X_{\text{CH}_4}=238.9$ppm.m and error $\pm 28.7$ppm.m. The error determined from the fit to the data can be interpreted as a sensitivity estimate for these experimental conditions. Here the high fringe visibility implies $\alpha\approx0.3$ and our LO based SNR theory does not strictly apply. However, balancing the longer integration times with the artificially lowered detector efficiency we might expect sensitivity better than 1ppm.m as predicted. Although we are an order of magnitude above this limit we assume this is due to technical noise such as vibrations and thermal drift over the experiment duration and expect this to reduce as we improve detector efficiency, data collection efficiency and apparatus stability.

\section{Conclusion}

The novel gas sensing method shown here enables a new generation of high-sensitivity methane sensors. The two-colour scheme provided by the technique makes it possible to illuminate the gas with a laser in the MIR region and detect light at SWIR. This allows the gas to interact with light at a wavelength for which it exhibits a high absorption while detecting modes at shorter wavelengths where low-noise and high-efficiency detectors are available. The experimental implementation proves that this scheme enables us to sense a very small methane concentration with high precision. Furthermore, it can be used to make a spectral analysis of the absorption lines of the gas sample. Calculations show that in high gain regimes $G>2.36\times10^{-4}$, the method has a sensitivity greater than the state-of-the-art IPDA and DIAL sensing techniques. In the future, this kind of nonlinear interferometer could be built in a compact fashion and light enough to be portable for field use. Analogous to IPDA LIDAR systems \cite{Titchener_2022}, this method can provide range measurements from the background scattering object by modulating the idler laser with a pseudo-random bit sequence to obtain the time of flight of the photons. High gains can be reached by using high-efficiency crystals, high-power pump laser (or pump build-up cavity) and tighter focussing into the NLC. This method can also be extended for sensing other gases as the signal and idler wavelengths can be tuned by the pump wavelength and the crystal phase matching.

\begin{backmatter}
\bmsection{Funding}
This work was funded by QuantIC - the UK Quantum Technology Hub in Quantum Imaging Grant No. 8031 EP/T00097X/1 and the Innovate UK Project 106174: SPLICE.

\bmsection{Acknowledgments}
The authors acknowledge Xiao Ai for the fruitful discussions with QLM Technology Ltd. and QuantIC and SPLICE projects for their support and encouragement.

\bmsection{Disclosures}
The authors declare no conflicts of interest.

\bmsection{Data availability} Data underlying the results presented in this paper are not publicly available at this time but may be obtained from the authors upon reasonable request.

\end{backmatter}



\end{document}